\documentclass[journal]{IEEEtran}
\usepackage{graphics}
\usepackage{mathrsfs}
\usepackage{color}
\usepackage{amsfonts} 
\usepackage{color} 
\usepackage{float}
\usepackage{cite}
\usepackage{graphicx}
\usepackage{amsmath}
\usepackage{amsthm}
\usepackage{multicol}
\usepackage{multirow}
\usepackage{color}
\usepackage{ifpdf}
\usepackage{flushend}
\usepackage{txfonts}
\usepackage{mdwlist}
\usepackage{caption}

\allowdisplaybreaks[4]

\usepackage{CJK}

\theoremstyle{definition} 
\theoremstyle{definition} 
\theoremstyle{definition} 
\theoremstyle{definition} 
\theoremstyle{assumption} 
\theoremstyle{definition} 
\theoremstyle{definition} 
\theoremstyle{definition} 
\theoremstyle{definition} 
\theoremstyle{definition} 

\usepackage{ifpdf}
\ifpdf
  \usepackage{epstopdf}
\fi

\newcommand{\R}{\mathbb{R}}

\definecolor{ttGreen}{rgb}{0.1,0.6,0.2}

\begin{document}

\title{Extremum Seeking Nonovershooting Control of Strict-Feedback Systems Under Unknown \\ Control Direction}

\author{Kaixin~Lu,~
        Ziliang~Lyu,~
        Yanfang~Mo,~
        Yiguang~Hong~
        and~Haoyong~Yu

\thanks{K. Lu (e-mail: kaixinlu@ln.edu.hk) is with the Department of Biomedical Engineering, National University of Singapore, Singapore and with the School of Data Science, Lingnan University, Hong Kong, China.}

\thanks{Z. Lyu (e-mail: ziliang.lyu@ntu.edu.sg) is with the School of Electrical and Electronic Engineering, Nanyang Technological University, Singapore.}

\thanks{Y. Mo (e-mail: yanfangmo@ln.edu.hk) is with the School of Data Science, Lingnan University, Hong Kong, China.}

\thanks{Y. Hong (e-mail: yghong@iss.ac.cn) is with the Department of Control Science and Engineering, Tongji University, Shanghai, China.}

\thanks{H. Yu (e-mail: bieyhy@nus.edu.sg) is with the Department of Biomedical Engineering, National University of Singapore, Singapore, and also with National University of Singapore (Suzhou) Research Institute, Suzhou, China.}
}

\maketitle

\begin{abstract}
This paper addresses the nonovershooting control problem for strict-feedback nonlinear systems with unknown control direction. We propose a method that integrates extremum seeking with Lie bracket-based design to achieve approximately nonovershooting tracking. The approach ensures that arbitrary reference trajectories can be tracked from below for any initial condition, with the overshoot reducible to arbitrarily small levels through parameter tuning. The method further provides a mechanism for enforcing high-relative-degree nonovershooting constraints in safety-critical scenarios involving unknown control directions.

\end{abstract}


\section{Introduction}
The problem of designing nonovershooting controllers is of great significance in practical applications. For example, in human-robot collaborative transportation tasks, the manipulator of a humanoid robot must move in a nonovershooting manner to prevent collisions with human coworkers. Another typical example is filling a tank to its desired level without overflow. Unlike conventional tracking control, nonovershooting control requires the system output to track a reference without exceeding it. Control design and analysis for linear systems with nonovershooting response have been well studied \cite{Vidyasagar_cdc_1990,Anderson_tacs1_2022}. For strict-feedback nonlinear systems, nonovershooting output tracking response was introduced and achieved in 2006 \cite{Krstic2006tac}. The nonovershooting controller in \cite{Krstic2006tac} can be interpreted as a safety filter, where the reference trajectory corresponds to the boundary of the safety set, and the first state-error variable effectively serves as a high-relative-degree control barrier function (CBF) \cite{xiangru_2015_ifac,Ames_2017_tac,tanxiao_2021_tac,Lyu_2023_tac,Lyu_2025_auto}. Then a mean-nonovershooting backstepping control method was proposed for stochastic systems \cite{wuquan_tac_2021}. Other developments in nonovershooting control were also addressed for multi-agent systems \cite{wuquan_auto_2025} and homogeneous systems \cite{Polyakov_tac_2023}.

Most existing nonovershooting control methods require the assumption of a known control direction, which refers to the motion direction of
a system under any control inputs. This assumption is important, since enforcing nonovershooting constraints requires the controller to act instantaneously in the correct direction. If the control direction is unknown, designing a nonovershooting controller is actually not trivial. Concretely, in nonlinear nonovershooting control (e.g., \cite{Krstic2006tac,wuquan_tac_2021,wuquan_auto_2025,Lyu_2025_auto}), the control direction is employed to derive the parameter gain selections that are used to prevent the overshoot. If the control direction is unknown, we may not find effective  gain selection to avoid the overshoot. Consequently, nonovershooting controller design for nonlinear systems under unknown control direction is a challenging problem.

The control problem under unknown control direction was posed by Morse \cite{Morse_1983} and first theoretically solved by Nussbaum \cite{Nussbaum_1983}. Nussbaum technique leverages oscillatory terms to adaptively identify the correct control direction (see in \cite{Xudong_tac_1998,ge_scl_2008,chenzy_auto_2019}). While it is a mainstream method currently, Nussbaum gain controllers may exhibit poor transient performance and repeated overshoots when the initial estimate of the control direction is incorrect, as pointed out by \cite{Scheinker2013tac}, where such issues were solved by proposing an elegant combination of extremum seeking (ES) and Lie bracket averaging theorem \cite{Gurvits1992ICRA}. Compared with the Nussbaum-type controllers, the ES controllers allow both the value and direction of control gain to be unknown and go through zero. In fact, ES is a real-time non-model-based optimization approach and is a powerful tool for finding the extremum value of an unknown map of dynamic systems. ES was first proposed in 1922 as an engineering invention \cite{Leblanc_1922} and sparked renewed interest in the field by the work \cite{Krs_es_2000}, which provided the first general proof of stability analysis by utilizing averaging and singular perturbation theory. Based on the breakthrough in \cite{Krs_es_2000}, ES has been applied to diverse areas including Nash equilibrium seeking \cite{Stankovic_cdc_2010}, stochastic control \cite{Stankovic_auto_2010}, PDE systems \cite{Oliveira_tac_2020} and safety filters \cite{Williams_2025ESSafe}. Using the notion of practical stability \cite{Moreau2000tac}, useful connections were established in \cite{Durr2013auto} between the optimal behavior of an ES system and the stability of its approximate system, referred to as the Lie bracket system, which provided a new way to solve regulation problems without prior system knowledge by seeking an extremum. Further works on ES and Lie bracket-based analysis were reported in \cite{Scheinker2014scl,Wang_tac_2024}.

Motivated by these observations, in this work, we aim to solve the nonovershooting output tracking problem by ES. Our objective is designing a controller such that the output $x_1(t)$ can track a given signal $y_r(t)$ from below, i.e., $$x_1(t)-y_r(t)\leq \bar{D},\quad \forall t\geq0$$ under unknown control direction. Here, $\bar{D}>0$ is an overshoot upper bound that can be made arbitrarily small. One main difficulty of solving this problem lies in designing a function to be minimized such that the resulting ES control action can dominate the overshoot. Although \cite{Scheinker2013tac} provided a systematic framework for constructing such a function, the associated ES controller may not necessarily guarantee the nonovershooting property. Moreover, we consider a trajectory tracking problem, which is much more difficult than stabilization, especially for systems with high relative degree\footnote{Relative degree is the minimal number of times the system output must be differentiated before the input appears explicitly.}. While a tracking extension was reported in \cite{Scheinker_cdc_2012} for systems with one relative degree, we consider nonlinear strict-feedback systems with higher relative degree, for which the ES-based tracking design is not trivial and needs new solutions. To the best of our knowledge, this problem remains open. The main contributions are as follows.
\begin{itemize}
    \item We propose an ES controller that achieve approximately nonovershooting tracking response under unknown control direction by minimizing a positive definite function. By Lie bracket approximation, we derive the conditions and parameter gain selections under which all the closed-loop signals are bounded and the overshoot can be dominated.

    \item The ES nonovershooting controller ensures that arbitrary reference trajectories can be tracked from below for any initial condition, with the overshoot reducible to arbitrarily small levels by systematically tuning design parameters. Moreover, the proposed ES controller provides a mechanism for enforcing high-relative-degree nonovershooting constraints in safety-critical scenarios involving unknown control directions.
\end{itemize}
\noindent
\textbf{Notation.} In this paper, $|\cdot|$ denotes the Euclidean norm. A function $\alpha:\R_{+}\rightarrow\R_{+}$ with $\alpha(0)=0$ is said to be of class $K$, if it is continuous and strictly increasing; it is of class $K_\infty$, if, in addition, $\alpha(s)\rightarrow\infty$ as $s\rightarrow\infty$.

\section{Problem Formulation and Preliminaries}

\subsection{Problem Formulation}
Consider a class of nonlinear strict-feedback systems:
\begin{align}\label{sys}
    \dot{x}_i & = x_{i+1} + \psi_i(\underline{x}_i), \quad i=1,...,n-1
    \nonumber \\
    \dot{x}_n & = g(\underline{x}_n)u + \psi_n(\underline{x}_n)
    \nonumber \\
    y & = x_1
\end{align}
where $x_i\in\R$ is the state and $\underline{x}_i=[x_1,...,x_i]^T\in\R^i$. $u\in\R$ is the control input and $y\in\R$ is the output. The function $\psi_i(\underline{x}_i)$ is known and is $n-1$ times differentiable, while the control gain $g(\underline{x}_n)$ is unknown, including both of its value $|g(\underline{x}_n)|$ and control direction sign$(g(\underline{x}_n))$.

\medskip
\noindent
\textbf{Problem of interests.} For a given signal $y_r(t)$, our objective is designing a controller $u$ for (\ref{sys}) under the unknown gain  $g(\underline{x}_n)$, such that the following results hold.
\begin{enumerate}
    \item All the closed-loop signals remain bounded.
    \item The tracking error can be controlled to an arbitrarily small residual around zero ultimately, i.e.,
    \begin{align}
    \lim_{t\rightarrow\infty} |x_1(t)-y_r(t)| \leq \iota(|d_1|)
         \label{hyt1}
    \end{align}
where $d_1$ is a nonzero constant and $\iota$ is a class $K_\infty$ function, both of which can be systematically tuned such that the value $\iota(|d_1|)$ is arbitrarily small.
\item The output $x_1$ tracks the reference $y_r$ from below, i.e.,
    \begin{align}
       x_1(t)-y_r(t) \leq \rho(|d_2|), \;\; \forall t\geq0\label{hyt2}
    \end{align}
where $d_2$ is a nonzero constant and $\rho$ is a class $K_\infty$ function, both of which can be systematically selected so that the value $\rho(|d_2|)$ is  as small as desired.
\end{enumerate}

\medskip
To facilitate the analysis, we make the following assumptions, where Assumption 1 is a regularity condition commonly used in backstepping design \cite{Krstic1995backstepping} and Assumption 2 is a standard requirement in classical works (see in \cite{Mudgett_tac_1985,Xudong_tac_1998}), which indicates that the control gain $g(\underline{x}_n)$ does not cross zero.

\medskip
\noindent
\textbf{Assumption 1.} The reference $y_r$ and its first $n$ times derivatives $\dot{y}_r$, $\ddot{y}_r$,...,$y_r^{(n)}$ are known, smooth and bounded.

\medskip
\noindent
\textbf{Assumption 2.} The control gain $g(\underline{x}_n)$ is uniformly bounded away from zero, i.e., $g(\underline{x}_n)^2\geq\xi_1$ for some constant $\xi_1>0$ and for all $\underline{x}_{n}\in\R^n$.

\medskip
\noindent
\textbf{Remark 1.} Compared with related works on nonovershooting control \cite{Krstic2006tac,wuquan_tac_2021,wuquan_auto_2025}, the control direction is unknown in our problem. Also note that while the works \cite{Scheinker2013tac,Scheinker2014scl} used ES and Lie bracket approximation to cope with unknown control direction, our new feature is twofold. First, in addition to the boundedness property (\ref{hyt1}), we further achieve the nonovershooting property (\ref{hyt2}). Second, we consider tracking problem, which is much more difficult, as pointed out by \cite{Scheinker_cdc_2012}. Although in \cite{Scheinker_cdc_2012}, a tracking extension was provided, it was for systems with relative degree one. In contrast, we address nonlinear strict-feedback systems with relative degree $n$, for which the controller design is substantially more challenging and requires new designs. \hfill $\blacksquare$

\medskip
One motivation of considering nonovershooting control under unknown control direction is that the results can be used to solve the safety filter design problem under unknown control direction. This problem has been a longstanding challenge in safety-critical control settings, as the appropriate direction in which unsafe control actions should be overridden is unknown.

\subsection{Practical Stability and Lie Bracket Approximation}

We first review the notion of practical stability \cite{Moreau2000tac}, which is closely related to Lyapunov stability and applies to systems depending on a parameter. Throughout this paper, we define this parameter as $\epsilon$.

We consider two systems: a system that depends on the parameter $\epsilon\in(0,\infty)$
\begin{align}\label{sys_mm_ocia}
    \dot{x} = f^\epsilon(t,x)
\end{align}
and a system:
\begin{align}\label{sys_mm_orgin}
    \dot{x} = f(t,x)
\end{align}
Let $\psi^{\epsilon}(t,t_0,x_0)$ be the solution of (\ref{sys_mm_ocia}) through $x(t_0)=x_0$, where the vector field $f^\epsilon:\R\times\R^n\rightarrow\R^n$ depends on $\epsilon$. Similarly, let $\psi(t,t_0,x_0)$ be the solution of (\ref{sys_mm_orgin}) through $x(t_0)=x_0$.

\medskip
\noindent
\textbf{Definition 1} (\emph{Converging Trajectories Property}, \cite{Moreau2000tac}) Systems (\ref{sys_mm_ocia}) and (\ref{sys_mm_orgin}) are said to satisfy the converging trajectories property, if for every $t_f\in(0,\infty)$ and compact set $\mathcal{C}\subseteq\R^n$ satisfying $\{(t,t_0,x_0)\in\R\times\R\times\R^n:t\in[t_0,t_0+t_f],x_0\in\mathcal{C}\}\subset \textup{Dom}\;\psi$, for every $d\in(0,\infty)$, there exists $\bar{\epsilon}>0$ such that for all $t_0\in\R$, and for all $x_0\in\mathcal{C}$, and for all $\epsilon\in(0,\bar{\epsilon})$
\begin{align}\label{converging_property}
    |\psi^\epsilon(t,t_0,x_0)-\psi(t,t_0,x_0)|<d, \quad \forall t\in[t_0,t_0+t_f]
\end{align}

The converging trajectories property (\ref{converging_property}) means that trajectory of (\ref{sys_mm_ocia}) converges uniformly on compact time intervals to trajectory of (\ref{sys_mm_orgin}) as $\epsilon\rightarrow0$.

\medskip
We define the form of stability for system (\ref{sys_mm_ocia}) as follows.

\medskip
\noindent
\textbf{Definition 2} (\emph{Semiglobal Practical Uniform Ultimate Boundedness With Bltimate Bound}, \emph{SPUUB}, \cite{Scheinker2014scl}) The origin of (\ref{sys_mm_ocia}) is SPUUB if there exists a constant $\delta>0$ such that the following conditions are satisfied
\begin{itemize}
    \item $(\epsilon,\delta)$-uniform stability.
    For every $c_2\in(\delta,\infty)$, there exist $c_1\in(\delta,\infty)$ and $\bar{\epsilon}\in(0,\infty)$ such that for all $t_0\in\R$ and for all $\epsilon\in(0,\bar{\epsilon})$
        \begin{align}\label{uniform_stable}
         |x_0|<c_1 \;\Rightarrow\; |\psi^\epsilon(t,t_0,x_0)|<c_2, \quad \forall t\in[t_0,\infty)
        \end{align}
    \item $(\epsilon,\delta)$-uniform ultimate boundedness.
    For every $c_1\in(0,\infty)$, there exist $c_2\in(\delta,\infty)$ and $\bar{\epsilon}\in(0,\infty)$ such that for all $t_0\in\R$ and for all $\epsilon\in(0,\bar{\epsilon})$
        \begin{align}\label{uniform_bounded}
         |x_0|<c_1 \;\Rightarrow\; |\psi^\epsilon(t,t_0,x_0)|<c_2, \quad \forall t\in[t_0,\infty)
        \end{align}
    \item $(\epsilon,\delta)$-global uniform attractivity.
    For every $c_1, c_2\in(\delta,\infty)$, there exist $t_f\in(0,\infty)$ and $\bar{\epsilon}\in(0,\infty)$ such that for all $t_0\in\R$ and for all $\epsilon\in(0,\bar{\epsilon})$
        \begin{align}\label{uniform_attractivity}
            |x_0|<c_1 \;\Rightarrow\; |\psi^\epsilon(t,t_0,x_0)|<c_2, \quad \forall t\in[t_0+t_f,\infty)
        \end{align}
\end{itemize}

If (\ref{uniform_stable})-(\ref{uniform_attractivity}) hold for all $\delta>0$, the origin of (\ref{sys_mm_ocia}) is semiglobally practically uniformly asymptotically stable (SPUAS), as introduced by \cite{Moreau2000tac}. If system (\ref{sys_mm_ocia}) does not depend on $\epsilon$, we omit the terms ``semi'' and ``practical'' in Definition 2, and the origin of (\ref{sys_mm_ocia})  is globally uniformly ultimately bounded with ultimate bound (GUUB).

\medskip
The following lemma links the stability properties of the systems in (\ref{sys_mm_ocia}) and (\ref{sys_mm_orgin}).

\medskip
\noindent
\textbf{Lemma 1} If systems (\ref{sys_mm_ocia}) and (\ref{sys_mm_orgin}) satisfy the property (\ref{converging_property}), and the origin of (\ref{sys_mm_orgin}) is GUUB, then the origin of (\ref{sys_mm_ocia}) is SPUUB.

\noindent
\textbf{Proof.} See the Appendix.

\medskip
Analogous to \cite{Moreau2000tac,Durr2013auto,Scheinker2013tac}, Lemma 1 only captures stability and not performance, and does not provide a systematic way for choosing $\bar{\epsilon}$. We only require the existence of $\bar{\epsilon}$ without explicitly considering a specific value. The choice of $\bar{\epsilon}$ usually depends on the set $\mathcal{C}$, the distance $d$ and the time $t_f$.

\medskip
Now, we review the preliminary on Lie bracket approximation. Consider a nonlinear system depending on $\epsilon\in(0,\infty)$
\begin{align}\label{sys_orign}
    \dot{x} = b_0(x) + \sum_{i=1}^mb_i(x)u_i^\epsilon(t,\theta)
\end{align}
where $x\in\R^n$, $b_i:\R^n\rightarrow\R^n$ for $i=0,...,m$ is smooth, $u_i^\epsilon(t,\theta)=\bar{u}_i(t)+\frac{1}{\sqrt{\epsilon}}\hat{u}_i(t,\theta)$ and $\theta=t/\epsilon$. $\hat{u}_i(t,\theta)$ is $T$-periodic in $T\in(0,\infty)$, i.e., $\hat{u}_i(t,\theta)=\hat{u}_i(t,\theta+T)$, and has zero average, i.e., $\int_0^T\hat{u}_i(t,\theta)d\theta=0$, for all $t,\theta\in \R$, $i=1,...,m$. Next, we consider a differential equation, which we call the Lie bracket system corresponding to (\ref{sys_orign})
\begin{align}\label{sys_Lie}
    \dot{z} = b_0(z)+ \sum_{i}^mb_i(z)\bar{u}_i(t) + \sum_{i<j}[b_i,b_j](z)v_{ji}(t), \;\;\; z(0)=x(0)
\end{align}
where
\begin{align}
   [b_i,b_j](z)  & = \frac{db_j(z)}{dz}b_i(z) - \frac{db_i(z)}{dz}b_j(z) \\
   v_{ji}(t)  & = \frac{1}{T}\int_0^T\hat{u}_j(t,\theta)\int_0^\theta \hat{u}_i(t,\tau)d\tau d\theta
\end{align}
\textbf{Lemma 2} \cite{Gurvits1992ICRA} Consider systems (\ref{sys_orign}) and (\ref{sys_Lie}). For period $T>0$ and $N=1,2,3,...$, there exists $\bar{\epsilon}>0$ such that for any $\epsilon\in(0,\bar{\epsilon})$, the trajectory $x(t)$ of (\ref{sys_orign}) is within a distance $\Delta(\epsilon)$ of the trajectory $z(t)$ of (\ref{sys_Lie}), that is,
\begin{align}
    \max_{t\in[0,NT]}\Big\{|x(t)-z(t)|\Big\}\leq \Delta(\epsilon)
\end{align}
where $\Delta(\epsilon)\rightarrow0$ as $\epsilon\rightarrow0$.

\medskip
\noindent
\textbf{Lemma 3} \cite{Scheinker2013tac} The solutions of (\ref{sys_orign}) and (\ref{sys_Lie}) satisfy the converging trajectories property (\ref{converging_property}).


\section{Main Results}

In this section, we first address the nonovershooting control problem for the case where $x_1(0) < y_r(0)$. We then address the case in which this condition is not necessarily satisfied. Finally, we provide a mechanism for enforcing high-relative-degree nonovershooting constraints in safety-critical scenarios involving unknown control directions.

Introduce a change of coordinates
\begin{align}
    h_1 & = x_1 - y_r \label{z1} \\
    h_i & = x_i - \alpha_{i-1} - y_r^{(i-1)}, \quad i=2,...,n \label{zi}
\end{align}
where the virtual controllers $\alpha_i$ are design as
\begin{align}
    \alpha_1 & = -c_1h_1 - \psi_1 \label{alpha1} \\
    \alpha_i & = -c_ih_i
     - \psi_i
    +\sum_{k=1}^{i-1}\Big[\frac{\partial\alpha_{i-1}}{\partial x_k}(x_{k+1}+\psi_k)
    + \frac{\partial\alpha_{i-1}}{\partial y_r^{(k-1)}}y_r^{(k)} \Big]\label{alphai}
\end{align}
where $c_i>0$ for $i=1,...,n-1$ are design parameters. Denote the vector $Y_r=[y_r,\dot{y}_r,...,y_r^{(n)}]^T$ and the function
\begin{align}
    \Psi(\underline{x}_n,Y_r)
    =\psi_n
    - \sum_{k=1}^{n-1}\Big[\frac{\partial\alpha_{i-1}}{\partial x_k}(x_{k+1}+\psi_k)
    + \frac{\partial\alpha_{i-1}}{\partial y_r^{(k-1)}}y_r^{(k)} \Big]
    - y_r^{(n)}
\end{align}
Since $\Psi(\underline{x}_n,Y_r)$ is a function of $\underline{x}_n$ and $Y_r$, with the definition of $h=[h_1,...,h_n]^T$ and Assumption 1, there exists a class $K_\infty$ function $\eta_1$ and a constant $\sigma_1>0$ such that
\begin{align}
    |\Psi(\underline{x}_n,Y_r)|\leq\eta_1(|h|)+\sigma_1 \label{Psi}
\end{align}
Design the control law
\begin{align}
    u =\alpha_{es}(h,t) =\sqrt{\omega}[\beta\cos(\omega t)-\lambda\sin(\omega t)V(h)]\label{u_nonovershoot}
\end{align}
where
\begin{align}
    V(h) & = \int_0^{|h|} r\eta_2(r)dr \label{Vz}
    \\
    \eta_2(r) & = \kappa_n \Big[\frac{c_n}{\kappa_n}+\Big(\eta_1(r)+\sigma_1\Big)
    +\Big(\eta_1(r)+\sigma_1\Big)^2\Big]\label{eta2}
\end{align}
with design parameters $c_n>0, \kappa_n>0$. With $V(h)$, we have
\begin{align}
\frac{\partial V(h)}{\partial h_n} &
    = |h|\eta_2(|h|)\frac{h_n}{|h|} = \eta_2(|h|)h_n
\end{align}
With (\ref{z1})-(\ref{zi}) and (\ref{alpha1})-(\ref{alphai}), the error dynamics of (\ref{sys}) is
\begin{align}
    \dot{h}_i & = -c_ih_i + h_{i+1}, \quad i=1,...,n-1 \label{dz1}
    \\
    \dot{h}_n & = gu + \psi_n - \dot{\alpha}_{n-1} - y_r^{(n)}\label{dzn}
\end{align}
With the controller (\ref{u_nonovershoot}), we get the Lie bracket averaging  dynamics of (\ref{dz1})-(\ref{dzn})
\begin{align}
    \dot{\bar{h}}_i & = -c_i\bar{h}_i + \bar{h}_{i+1}, \quad i=1,...,n-1
    \label{d:bar:z1}
    \\
    \dot{\bar{h}}_n & = \Psi(\bar{\underline{x}}_n,Y_r)
    -\lambda\beta g^2(\bar{\underline{x}}_n)
    \frac{\partial V(\bar{h})}{\partial \bar{h}_n}
    \nonumber \\
    & = \Psi(\bar{\underline{x}}_n,Y_r)
    -\lambda\beta g^2\eta_2(|\bar{h}|)\bar{h}_n \label{d:bar:zn}
\end{align}
with $\bar{h}(0)=h(0)$. Denoting $\epsilon=1/\omega$ and $\bar{\epsilon}=\underline{\omega}$, and by Lemma 2 and Lemma 3, there exists $\underline{\omega}$ such that for all $\omega\in(\underline{\omega},\infty)$, the trajectories of $h(t)$ and $\bar{h}(t)$ satisfy
\begin{align}\label{z-barz}
    |h(t)-\bar{h}(t)|<\Delta(1/\omega), \quad \forall t\in[0,t_f)
\end{align}
for every $t_f\in(0,\infty)$, where $\Delta(1/\omega)$ can be made arbitrarily small by increasing $\omega$, and $\Delta(1/\omega)\rightarrow0$ as $\omega\rightarrow\infty$.

\medskip
\noindent
\textbf{Remark 2.} Different from standard backstepping design \cite{Krstic1995backstepping}, we omit the term $-h_{i-1}$ in the design of $\alpha_i$ for $i=2,...,n-1$ and the resulting error dynamics is (\ref{dz1})-(\ref{dzn}), which is necessary to achieve the nonovershooting property and is inspired by \cite{Krstic2006tac}. In current works on nonovershooting control (e.g., \cite{Krstic2006tac,wuquan_tac_2021,wuquan_auto_2025}), the control gain $g(\underline{x}_n)$ is necessary for nonovershooting design, as it is used to derive the gain selections for the parameters $c_i$ that are employed to prevent overshoot. If $g(\underline{x}_n)$ is unknown, we cannot find effective selection of $c_i$ to avoid the overshoot.

We use ES to solve this problem. Unlike traditional control, in an ES scheme, one cannot directly construct the controller to dominate the unstable terms or overshoot. Instead, these terms need to be canceled by the gradient of the function designed in the ES controller, which is therefore more challenging, particularly for tracking problems with high relative degree. To overcome the difficulty, we propose an ES nonovershooting controller (\ref{u_nonovershoot}), which is a minimum-seeking controller for a positive definite function $V(h)$ designed in (\ref{Vz}). Under (\ref{u_nonovershoot}), we compute the Lie bracket average system (\ref{d:bar:z1})-(\ref{d:bar:zn}). The ES scheme introduces a gradient of $V(\bar{h})$, which can dominate the potential overshoot and stabilize the closed-loop average system.

In the following theorem, we provide new conditions and gain selections on parameters $c_i$ under which the gradient of $V(\bar{h})$ and its related term can dominate the unstable terms and overshoot such that the average system is GUUB. Stability and nonovershooting property of the closed-loop system (\ref{sys}), (\ref{u_nonovershoot}) are established via the connections between the Lie bracket average system and the original system provided in Lemma 1.

\hfill $\blacksquare$

The main results are summarized in Theorem 1.

\medskip
\noindent
\textbf{Theorem 1} Consider the closed-loop system (\ref{sys}), (\ref{u_nonovershoot}). If $c_i$ are chosen such that $c_i>1$ for $i=1,...,n$, and $\beta$ and $\lambda$ are chosen such that
\begin{align}
    \lambda\beta\geq1/{\xi_1} \label{lambdabeta}
\end{align}
then the following results hold.
\begin{itemize}
  \item[\textbf{i)}] The equilibrium $h=0$ of the closed-loop system (\ref{sys}), (\ref{u_nonovershoot}) is SPUUB.
  \item[\textbf{ii)}] The tracking error $h_1(t)=x_1(t)-y_r(t)$ ultimately  converges to an arbitrarily small residual around zero, i.e.,
        \begin{align}
            \lim_{t\rightarrow\infty} |h_1(t)| \leq \iota(|d_1|) \label{z1:leq:D:inifity}
        \end{align}
where $\iota(s)=s$ is a $K_\infty$ function,  $d_1=\sqrt{1/(\kappa_n(c_m-1))} +\Delta(1/\omega)$ is a constant. Here, $c_m=\min\{c_i, \ i=1,...,n\}$. The value $\iota(|d_1|)$ can be reduced as small as desired by increasing the parameters $\kappa_n$, $c_i$ and the frequency $\omega$.

  \item[\textbf{iii)}] If $h_1(0)<0$, then the choice $c_i>
\max\{\underline{c}_i,1\}$, where
\begin{align}
    \underline{c}_i & = -\frac{1}{h_{i}(0)}\Big\{x_{i+1}(0)+\psi_{i}(\underline{x}_{i}(0))-y_r^{(i)}(0)
    \nonumber \\
    & \quad-\sum_{k=1}^{i-1}\Big[
    \frac{\partial\alpha_{i-1}(\underline{x}_{i-1}(0),y_r(0),...,y_r^{(i-1)}(0))}{\partial y_r^{(k-1)}(0)}y_r^{(k)}(0)
    \nonumber \\
    & \quad
    +
    \frac{\partial\alpha_{i-1}
    (\underline{x}_{i-1}(0),y_r(0),...,y_r^{(i-1)}(0))}{\partial x_k}
    \nonumber \\
    &\quad \times
    \Big(x_{k+1}(0)
    +\psi_k(\underline{x}_k(0))\Big)\Big] \Big\}\label{ci_underline}
\end{align}
for $i=1,...,n-1$ and $c_n>1$ guarantees that
\begin{align}
    h_1(t) \leq \rho(|d_2|), \quad \forall t\geq0 \label{z1:leq:D}
\end{align}
Here, $\rho(s)=s$ is a $K_\infty$ function, and $d_2=1/(\kappa_n\prod_{i=1}^nc_i)+\Delta(1/\omega)$. The value $\rho(|d_2|)$ can be made arbitrarily small by increasing the parameters $\kappa_n$, $c_i$ and frequency $\omega$.
\end{itemize}

\noindent
\textbf{Proof.} \textbf{i)}  Denote $\mu_n(|\bar{h}|)=\lambda\beta g^2
    \eta_2(|\bar{h}|)$ and consider a Lyapunov function candidate $W(\bar{h})=\frac{1}{2}\bar{h}^T\bar{h}$. By (\ref{Psi}), (\ref{lambdabeta}), and Assumption 1, we have $\mu_n\geq\kappa_n|\Psi|$ and
\begin{align}\label{dWz}
    \dot{W}(\bar{h}) & = -\sum_{i=1}^{n-1}c_i\bar{h}_i^2 +\sum_{i=1}^{n-1}\bar{h}_i\bar{h}_{i+1}
    +\Psi \bar{h}_n - \mu_n\bar{h}_n^2
    \nonumber \\
    & \leq -\sum_{i=1}^{n}(c_i-1)\bar{h}_i^2
    +\Big(\eta_1(|\bar{h}|)+\sigma_1\Big)|\bar{h}_n|
    \nonumber \\
    & \quad
    -\kappa_n\Big(\eta_1(|\bar{h}|)+\sigma_1\Big)\bar{h}_n^2
    -\kappa_n\Big(\eta_1(|\bar{h}|)+\sigma_1\Big)^2\bar{h}_n^2
    \nonumber \\
    & \leq -2(c_{m}-1)W(\bar{h})+ \frac{1}{\kappa_n}
\end{align}
Therefore, the averaged system (\ref{d:bar:z1})-(\ref{d:bar:zn}) is uniformly  ultimately bounded with ultimate bound $\sqrt{1/(\kappa_n(c_m-1))}$, which can be made sufficiently small by increasing $\kappa_n$ and $c_i$. Because the closed-loop system (\ref{sys}), (\ref{u_nonovershoot}) and the averaged system (\ref{d:bar:z1})-(\ref{d:bar:zn}) satisfy the tracking trajectories property, by Lemma 1, the equilibrium $h=0$ of the closed-loop system (\ref{sys}), (\ref{u_nonovershoot}) is SPUUB.

\textbf{ii)} By (\ref{dWz}), $\lim_{t\rightarrow\infty} |\bar{h}_1(t)| \leq\sqrt{1/(\kappa_n(c_m-1))}$. Then by Lemma 1 and (\ref{z-barz}), we have (\ref{z1:leq:D:inifity}).

\textbf{iii)} Applying the variation of constant formula to (\ref{d:bar:zn}) we get
\begin{align}
    \bar{h}_n(t) & =
    \bar{h}_n(0)e^{-\int_0^t\mu_n(\tau)d\tau}
    + \int_0^t e^{-\int_s^t\mu_n(\tau)d\tau}\Psi(s)ds
    \nonumber \\
    & \leq \bar{h}_n(0)e^{-\int_0^t\mu_n(\tau)d\tau}
    + \frac{1}{\kappa_n}\int_0^te^{-\int_s^t\mu_n(\tau)d\tau}\mu_n(s)ds
    \nonumber \\
    & \leq \bar{h}_n(0)e^{-\int_0^t\mu_n(\tau)d\tau}
    \nonumber \\
    & \quad
    + \frac{1}{\kappa_n}e^{-\int_0^t\mu_n(\tau)d\tau}
    \int_0^te^{\int_0^s\mu_n(\tau)d\tau}\mu_n(s)ds
    \nonumber \\
    & = \bar{h}_n(0)e^{-\int_0^t\mu_n(\tau)d\tau}
    \nonumber \\
    & \quad + \frac{1}{\kappa_n}e^{-\int_0^t\mu_n(\tau)d\tau}
    \int_0^te^{\int_0^s\mu_n(\tau)d\tau}d\Big(\int_0^s\mu_n(\tau)d\tau\Big)
    \nonumber \\
    & \leq \bar{h}_n(0)e^{-\int_0^t\mu_n(\tau)d\tau} + \frac{1}{\kappa_n}
\end{align}
Similarly, by induction, we have
\begin{align}\label{zn-1}
    \bar{h}_{n-1}(t) & = \bar{h}_{n-1}(0)e^{-c_{n-1}t}
    + \int_0^t e^{-c_{n-1}(t-s)}\bar{h}_n(s)ds
    \nonumber \\
    & \leq \bar{h}_{n-1}(0)e^{-c_{n-1}t}
    + \bar{h}_n(0)\int_0^t e^{-c_{n-1}(t-s)-\int_0^s\mu_n(\tau)d\tau}ds \nonumber \\
    & \quad + \frac{1}{\kappa_n}\int_0^t e^{-c_{n-1}(t-s)}ds \nonumber \\
    & \leq \varphi_{n-1}(t)
    + \frac{1}{c_{n-1}\kappa_n}
\end{align}
where $\varphi_{n-1}(t)=\bar{h}_{n-1}(0)e^{-c_{n-1}t}
    + \bar{h}_n(0)\int_0^t e^{-c_{n-1}(t-s)-\int_0^s\mu_n(\tau)d\tau}ds$. With the choice $c_i>\max\{\underline{c}_i,1\}$ and $h_1(0)<0$, the initialization of the state error variable $h_i(0)<0$ holds for $i=2,...n$. Due to $\bar{h}(0)=h(0)$, the function $\varphi_{n-1}(t)$ is not positive since $\bar{h}_n(0)$ and $\bar{h}_{n-1}(0)$ are negative. Similarly, we have
\begin{align}\label{barz1}
    \bar{h}_1(t)\leq \varphi_1(t) + \frac{1}{\kappa_n\prod_{i=1}^nc_i}
\end{align}
where $\varphi_1(t)$ is not positive. With Lemma 1, (\ref{z-barz}) and (\ref{barz1}), we have (\ref{z1:leq:D}). The proof is completed. \hfill $\blacksquare$

\medskip
Theorem 1 indicates that if we properly choose the parameter $c_i$ such that $h_i(0)<0$, the nonovershooting property (\ref{z1:leq:D}) is achieved. Next, we show that if such initial condition is not satisfied, (\ref{z1:leq:D}) still holds by a new selection of $c_i$.

\medskip
\noindent
\textbf{Theorem 2} If $h_1(0)<0$ does not hold, then the choice
\begin{align}\label{newchoice}
    c_1>c_2>\cdots>c_n>1
\end{align}
renders approximately nonovershooting tracking of system (\ref{sys}) with a larger overshoot bound than that in (\ref{z1:leq:D}), i.e.,
\begin{align}
    h_1(t) \leq \rho(|d_2|)
    + \sum_{i=1}^na_i|h_i(0)|e^{-c_it}, \quad \forall t\geq0
     \label{z1:leq:D_2}
\end{align}
where $a_1=1$ and $a_i=1/\prod_{k=1}^{i-1}(c_k-c_i)$ for $i=2,...,n$, are positive constants.

\noindent
\textbf{Proof.} By (\ref{zn-1}) and a tedious induction, we have
\begin{align}
    \bar{h}_1(t) \leq \frac{1}{\kappa_n\prod_{i=1}^nc_i}  + \sum_{i=1}^n\bar{h}_i(0)\phi_i(t) \label{barz1:leq:D_2}
\end{align}
where $\phi_1(t)=e^{-c_1t}$, $\phi_2(t)=\int_0^te^{-c_1(t-s_1)-c_2s_1}ds_1$,
\begin{align}
   \phi_i(t) & = \int_0^te^{-c_1(t-s_1)}\int_0^{s_1}e^{-c_2(s_1-s_2)}
   \cdots
    \int_0^{s_{i-3}}e^{-c_{i-2}(s_{i-3}-s_{i-2})}
   \nonumber \\
    & \quad \int_{0}^{s_{i-2}}e^{-c_{i-1}(s_{i-2}-s_{i-1})
    -c_is_{i-1}}d{s_{i-1}}d{s_{i-2}}\cdots d{s_1}
    \\
    \phi_n(t) & = \int_0^te^{-c_1(t-s_1)}\int_0^{s_1}e^{-c_2(s_1-s_2)}\cdots
    \int_0^{s_{n-3}}e^{-c_{n-2}(s_{n-3}-s_{n-2})}
    \nonumber \\
    & \quad
    \int_{0}^{s_{n-2}}e^{-c_{n-1}(s_{n-2}-s_{n-1}) -\int_0^{s_{n-1}}\mu_n(\tau)d\tau}d{s_{n-1}}d{s_{n-2}}\cdots d{s_1}
\end{align}
We calculate the innermost integral of $\phi_n(t)$ and then proceed outward. With (\ref{eta2}), (\ref{lambdabeta}), (\ref{newchoice}) and the fact that $\mu_n(|\bar{h}|)=\lambda\beta g^2\eta_2(|\bar{h}|)$, we have $\mu_n(|\bar{h}|)\geq c_n$. Then
\begin{align}\label{sn-2}
   & \quad
   \int_{0}^{s_{n-2}}e^{-c_{n-1}(s_{n-2}-s_{n-1}) -\int_0^{s_{n-1}}\mu_n(\tau)d\tau}d{s_{n-1}}
   \nonumber \\
   & \leq
   \int_{0}^{s_{n-2}}e^{-c_{n-1}(s_{n-2}-s_{n-1}) -c_ns_{n-1}}d{s_{n-1}}
   \nonumber \\
   & = \frac{1}{c_{n-1}-c_n}(e^{-c_ns_{n-2}}-e^{-c_{n-1}s_{n-2}})
   \leq \frac{1}{c_{n-1}-c_n}e^{-c_ns_{n-2}}
\end{align}
Calculate the second innermost integral of $\phi_n(t)$ along with the result (\ref{sn-2}) yielding that
\begin{align}
   & \quad \frac{1}{c_{n-1}-c_n}\int_0^{s_{n-3}}e^{-c_{n-2}(s_{n-3}-s_{n-2})-c_ns_{n-2}}ds_{n-2}
    \nonumber \\
   & = \frac{1}{(c_{n-1}-c_n)(c_{n-2}-c_n)}(e^{-c_ns_{n-3}}-e^{-c_{n-2}s_{n-3}})
   \nonumber \\
   & \leq \frac{1}{(c_{n-1}-c_n)(c_{n-2}-c_n)}e^{-c_ns_{n-3}}
\end{align}
Then, by induction,
\begin{align}
    \phi_n(t)\leq\frac{1}{\prod_{i=1}^{n-1}(c_i-c_n)}e^{-c_nt}
\end{align}
With the same method, we have $\phi_i(t)\leq\frac{1}{\prod_{k=1}^{i-1}(c_k-c_i)}e^{-c_it}$. By (\ref{z-barz}) and $\bar{h}_i(0)=h_i(0)$, we have
\begin{align}
    h_1(t) \leq \frac{1}{\kappa_n\prod_{i=1}^nc_i} + \Delta(1/\omega) + |h_1(0)|e^{-c_1t}
    + \sum_{i=2}^n\frac{|h_i(0)|e^{-c_it}}{\prod_{k=1}^{i-1}(c_k-c_i)}
    \nonumber 
\end{align}
which equals to (\ref{z1:leq:D_2}). The proof is completed. \hfill $\blacksquare$

\medskip
\noindent
\textbf{Remark 3.} The gain selection (\ref{ci_underline}) requires the initialization $h_1(0)<0$, because if $h_1(0)=0$, we cannot choose a valid $c_i$; if $h_1(0)>0$, there may arise an overshoot that cannot be arbitrarily reduced (e.g., recalling (\ref{barz1}), $\varphi_1(t)<0$ may not hold when $h_1(0)>0$ and (\ref{z1:leq:D}) may not be achieved). In Theorem 2, we provide a new selection on $c_i$ to solve this problem such that the ES nonovershooting controller is effective under any initial conditions. \hfill $\blacksquare$

\medskip
Next, we show that the nonovershooting controller $u=\alpha_{es}$ in (\ref{u_nonovershoot}) can be employed to produce a safety-critical controller under nonovershooting constraint $\mathcal{H}(x_1,t)=y_r(t)-x_1(t)\geq0$. The safe controller is
\begin{align}\label{safe_controller}
    u(t) =
    \begin{cases}
        u_0,         & \mathcal{H}(x_1,t)\geq0 \\
        \alpha_{es}, & \textup{otherwise}
    \end{cases}
\end{align}
where $u_0$ is a nominal controller. The safe controller (\ref{safe_controller}) applies the user's nominal control $u_0$ when the system operates in the safe set $\mathcal{H}(x_1,t)\geq0$ and overrides $u_0$ with the ES nonovershooting controller $\alpha_{es}$ in the domain where the user's nominal control commands operation in the unsafe set. Such a design is motivated by the safe learning process (see in \cite{Allgower_2007,Akametalu_2018}). Note that at any switching instant $t_k$ of the safe controller (\ref{safe_controller}), the new initialization $h_1(t_k)<0$ required in Theorem 1 may not hold. Theorem 2 can  solve this problem and the safe controller (\ref{safe_controller}) is thus valid under any initial conditions.

\section{Illustrative Examples}

\begin{figure}
\centering
\includegraphics[width = 7.9cm]{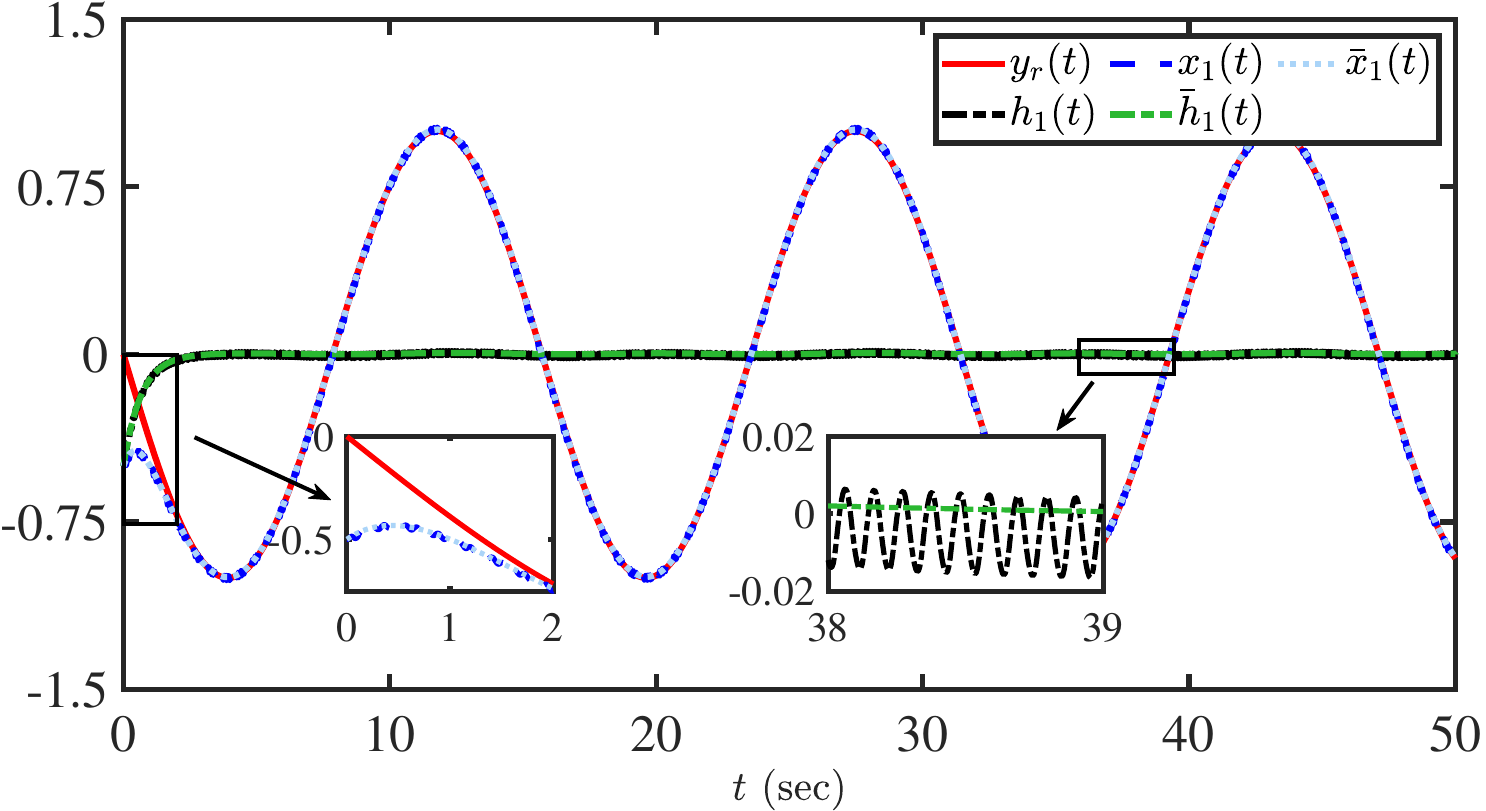}
\caption{Tracking trajectories under the ES nonovershooting controller.}\label{figure1_tra}
\end{figure}

\begin{figure}
\centering
\includegraphics[width = 7.5cm]{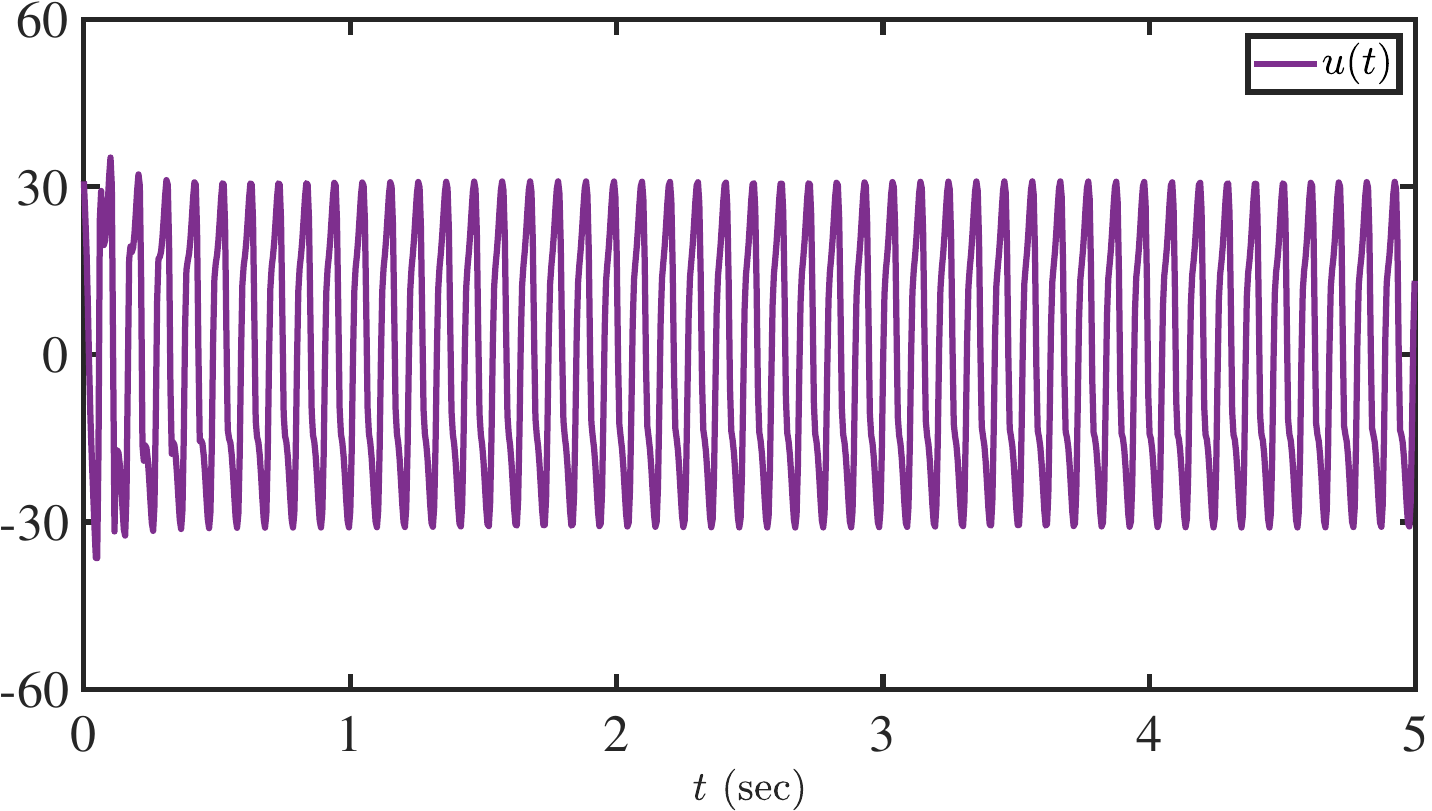}
\caption{Evolutions of the ES nonovershooting controller.}\label{figure6_u}
\end{figure}

In this section, we consider a nonlinear strict-feedback system
\begin{align}\label{ex_sys}
    \dot{x}_1 = x_2, \quad
    \dot{x}_2 = (0.2\sin x_2 + 1.2)u + x_1^2
\end{align}
The system functions $\psi_1(\underline{x}_1)=0$ and $\psi_2(\underline{x}_2)=x_1^2$ are known for control design, while the control gain $g(\underline{x}_2)=0.2\sin x_2 + 1.2$ is unknown for control design. The control objective is making the output $x_1$ track the reference $y_r=-\sin (0.4t)$ from below, i.e., the tracking error $h_1(t)=x_1-y_r\leq d_2$ for all $t\geq0$, where the overshoot upper bound $d_2$ can be made arbitrarily small. Choose $c_1=2$, $c_2=1.5$, $\kappa_2=1.1$, $\omega=60$, $\lambda=4$ and $\beta=0.8$. The initialization is $x(0)=[x_1(0),x_2(0)]^T=[-0.5,0]^T$. As is evident from Fig.~\ref{figure1_tra}, the output $x_1$ tracks the reference $y_r$ from below approximately all the time. Evolutions of the nonovershooting controller are shown in Fig.~\ref{figure6_u}. As the controller exhibits oscillatory behavior, we zoom in on the interval $[0,5]$s (instead of the full $[0,50]$s) to clearly illustrate its evolution.

\begin{figure}
\centering
\includegraphics[width = 7.5cm]{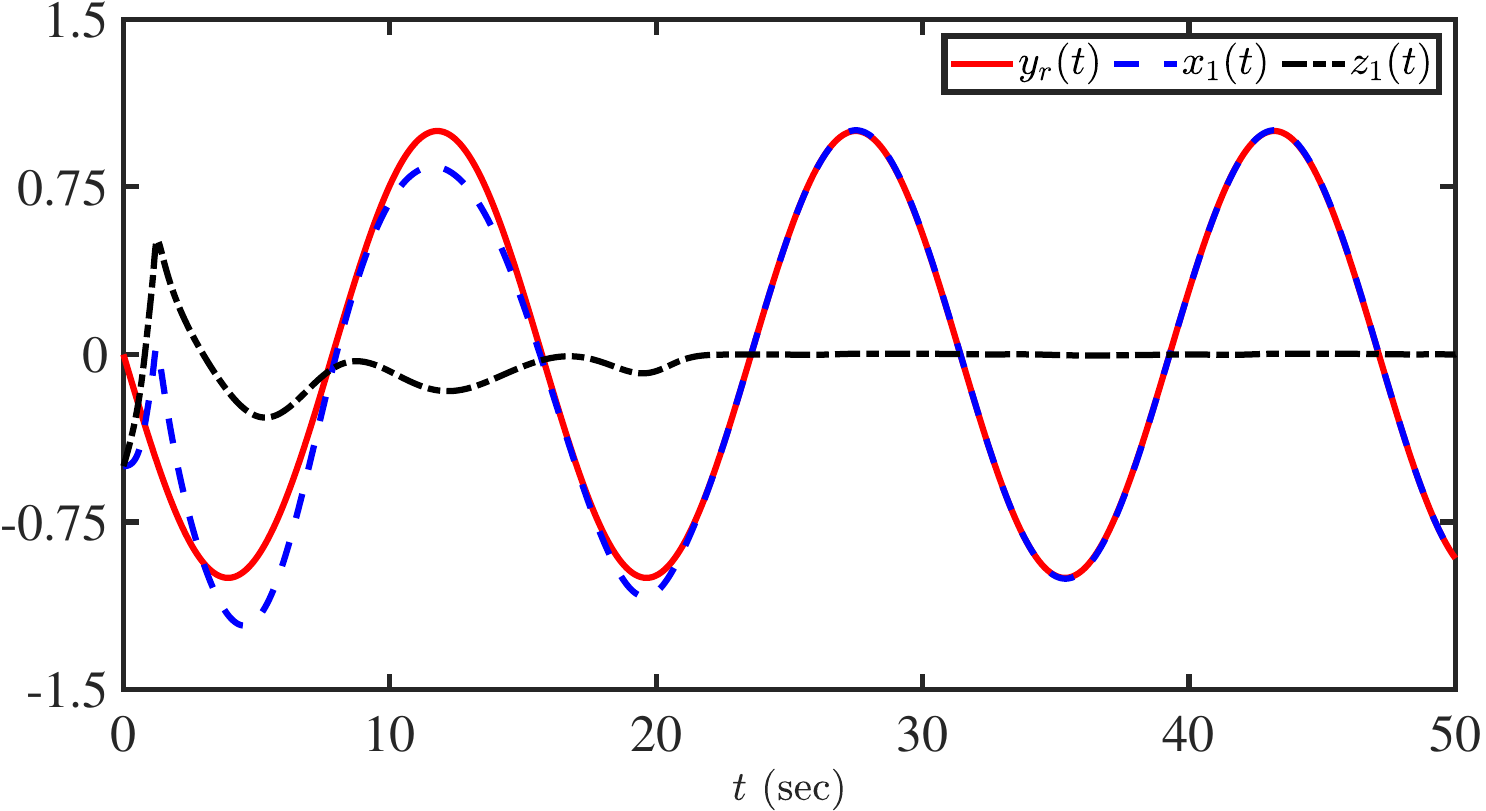}
\caption{Tracking trajectories under the Nussbaum controller.}\label{figure2_nussbuam}
\end{figure}

We also compare our controller to the controller designed by the Nussbaum technique:
\begin{align}\label{nuss_control}
    u = \vartheta^2\cos(\vartheta)\alpha_2, \;\;\dot{\vartheta}=-\alpha_2z_2
\end{align}
where $\alpha_2$ is designed by backstepping \cite{Krstic1995backstepping} with the same parameters $c_1$ and $c_2$ as in our controller, $z_1$ and $z_2$ are the corresponding state error variables. As shown in Fig.~\ref{figure2_nussbuam}, a very large overshoot is induced, which is caused by a wrong initial guess of the control direction, as indicated by \cite{Scheinker2013tac}. We note that although the Nussbaum controller (\ref{nuss_control}) causes a large overshoot, it can achieve asymptotic tracking. In contrast, our method provides approximately nonovershooting tracking response but guarantees only SPUUB performance.

\begin{figure}[H]
\centering
\includegraphics[width = 7.5cm]{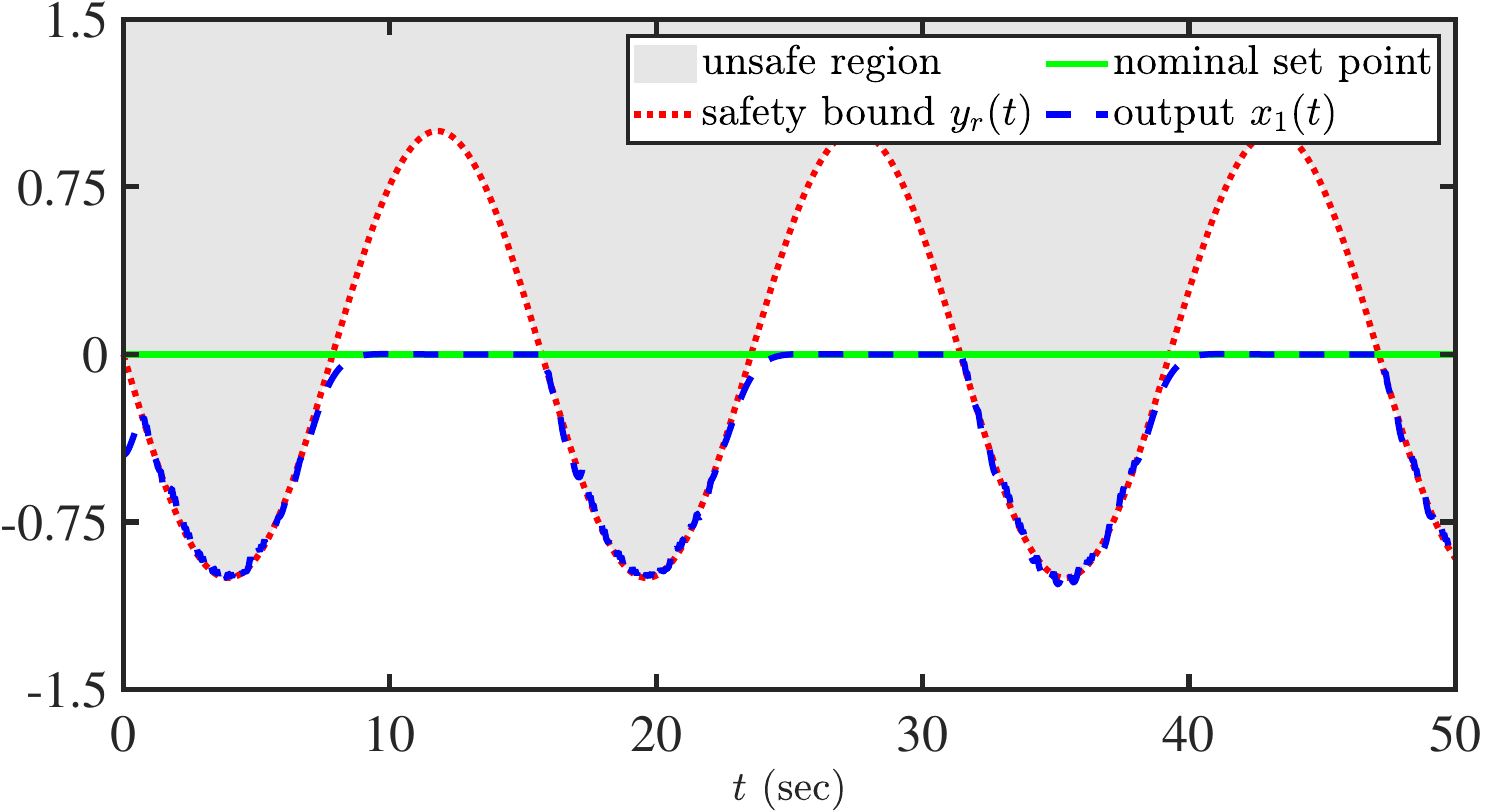}
\caption{Safe regulation trajectories with safe initialization.}\label{figure4_ride_setpoint}
\end{figure}

Next, we show that the proposed nonovershooting controller can be used to design a safety filter under nonovershooting constraints. Consider again the system (\ref{ex_sys}). The control objective is regulating the output $x_1$ to zero, while ensuring that the safety constraint $\mathcal{H}(x_1,t)=y_r(t)-x_1(t)\geq0$ is satisfied for all $t\geq0$, where $y_r(t)=-\sin(0.4t)$. The nominal controller $u_0$ is designed using backstepping \cite{Krstic1995backstepping}. To avoid unnecessary complexity, we assume that $u_0$ knows the function $g(\underline{x}_2)$. All the parameters are the same as previously, and the system is initialized with $x(0)=[-0.45,0]^T$. As shown in Fig.~\ref{figure4_ride_setpoint}, the safe controller (\ref{safe_controller}) checks whether the output $x_1$ lies inside the safe set $\mathcal{H}(x_1,t)\geq0$. If not, the system is steered toward and along the boundary of the safe set $y_r$ by $u=\alpha_{es}$; if so, it is regulated to zero by $u=u_0$ directly.

\begin{figure}[H]
\centering
\includegraphics[width = 7.5cm]{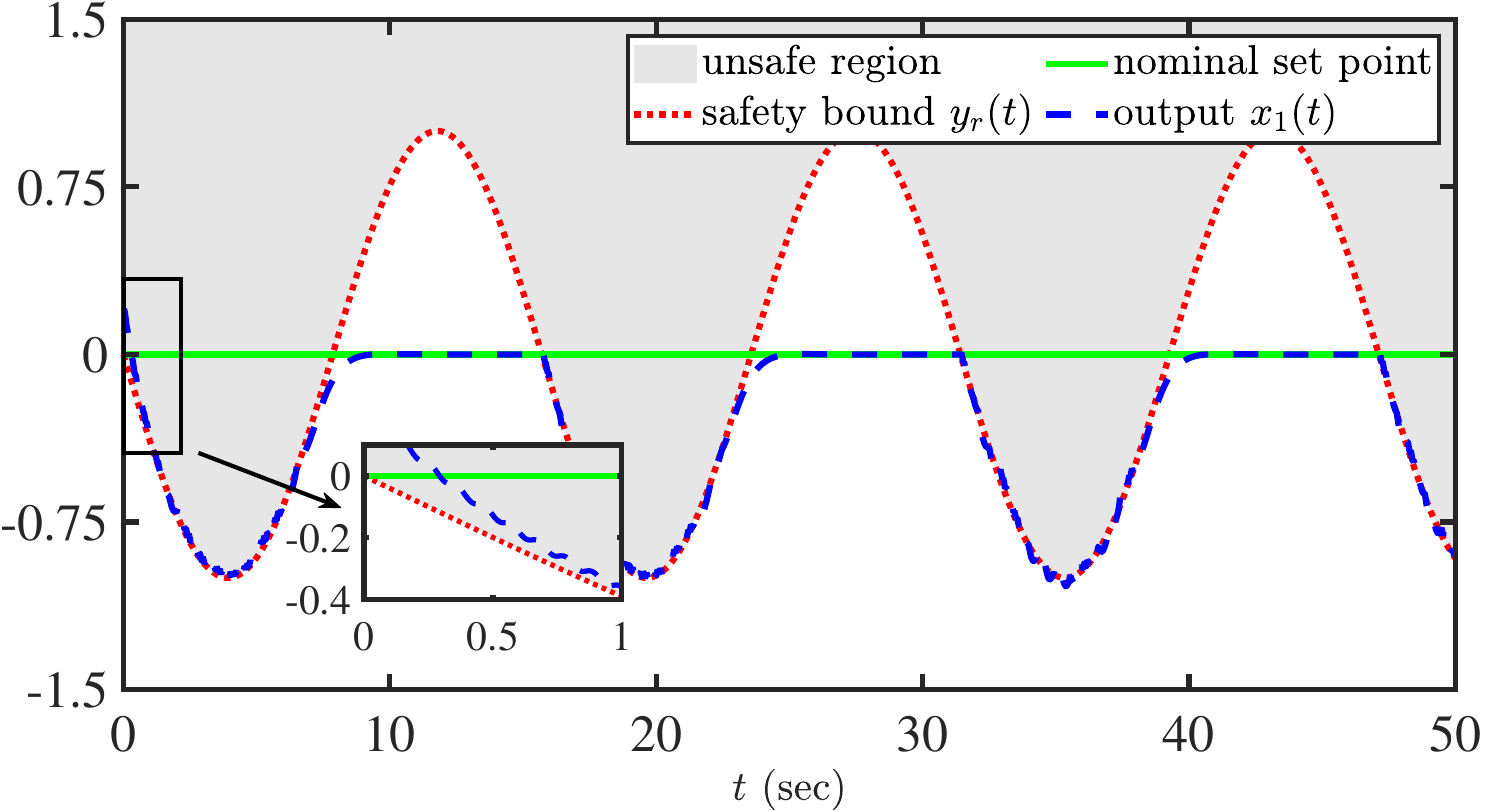}
\caption{Safe regulation trajectories with unsafe initialization.}\label{figure5_ride_setpoint}
\end{figure}

In Theorem 2, we show that even if the condition $h_1(0)<0$ does not hold, by selecting $c_1>c_2>1$, the nonovershooting property still holds. We now validate this result. Since the selection $c_1=2$ and $c_2=1.5$ naturally satisfies $c_1>c_2>1$, we do not need to reselect the parameters. Choose $x(0)=[0.2,0]^T$, which yields $h_1(0)=0.2>0$ and $h_2(0)=0.8>0$. Similar results are obtained from Fig.~\ref{figure5_ride_setpoint}. At the beginning, the output is in the unsafe region and thus, it is steered toward and along the boundary of the safe set by $u=\alpha_{es}$ first rather than regulated to zero.

\section{Conclusion}

In this work, an ES nonovershooting controller was proposed for strict-feedback systems under unknown control direction. With the proposed method, any time-varying reference trajectories could be tracked from below approximately for any initial conditions, where the upper bound of the overshoot could be made arbitrarily small by parameter tuning. The proposed nonovershooting controller provided a notable advantage, as it could be directly employed as an override controller to handle high-relative-degree nonovershooting constraints under unknown control direction. Simulations validated the proposed results.


\ifCLASSOPTIONcaptionsoff
  \newpage
\fi


\begin{thebibliography}{10}
\bibitem{Anderson_tacs1_2022}
B.~D. Anderson, M.~Deistler, L.~Farina, and L.~Benvenuti,
``Nonnegative realization of a linear system with nonnegative impulse response,''
{\em IEEE Transactions on Circuits and Systems I: Fundamental Theory and Applications},
vol.~43, no.~2, pp.~134--142, 2002.

\bibitem{Vidyasagar_cdc_1990}
G.~Deodhare and M.~Vidyasagar,
``Design of non-overshooting feedback control systems,''
in {\em 29th IEEE Conference on Decision and Control},
pp.~1827--1834, IEEE, 1990.


\bibitem{Krstic2006tac}
M.~Krsti\'c and M.~Bement,
``Nonovershooting control of strict-feedback nonlinear systems,''
{\em IEEE Transactions on Automatic Control},
vol.~51, no.~12, pp.~1938--1943, 2006.

\bibitem{xiangru_2015_ifac}
X.~Xu, P.~Tabuada, J.~W.~Grizzle, and A.~D.~Ames,
``Robustness of control barrier functions for safety critical control,''
{\em IFAC-PapersOnLine},
vol.~48, no.~27, pp.~54--61, 2015.

\bibitem{Ames_2017_tac}
A.~D.~Ames, X.~Xu, J.~W.~Grizzle, and P.~Tabuada,
``Control barrier function based quadratic programs for safety critical systems,''
{\em IEEE Transactions on Automatic Control},
vol.~62, no.~8, pp.~3861--3876, 2016.

\bibitem{tanxiao_2021_tac}
X.~Tan, W.~S.~Cortez, and D.~V.~Dimarogonas,
``High-order barrier functions: Robustness, safety, and performance-critical control,''
{\em IEEE Transactions on Automatic Control},
vol.~67, no.~6, pp.~3021--3028, 2021.

\bibitem{Lyu_2023_tac}
Z.~Lyu, X.~Xu, and Y.~Hong,
``Small-gain theorem for safety verification under high-relative-degree constraints,''
{\em IEEE Transactions on Automatic Control},
vol.~69, no.~6, pp.~3717--3731, 2023.

\bibitem{Lyu_2025_auto}
Z.~Lyu, M.~Krsti\'c, K.~Lu, Y.~Hong, and L.~Xie,
``Adaptive override control under high-relative-degree nonovershooting constraints,''
{\em arXiv preprint arXiv:2509.18988}, 2025.

\bibitem{wuquan_tac_2021}
W.~Li and M.~Krsti\'c,
``Mean-nonovershooting control of stochastic nonlinear systems,''
{\em IEEE Transactions on Automatic Control},
vol.~66, no.~12, pp.~5756--5771, 2021.

\bibitem{wuquan_auto_2025}
W.~Li, H.~Wang, L.~Liu, and G.~Feng,
``Distributed mean-nonovershooting control of stochastic nonlinear multi-agent systems,''
{\em Automatica},
vol.~179, art.~no.~112397, 2025.

\bibitem{Polyakov_tac_2023}
A.~Polyakov and M.~Krsti\'c,
``Finite- and fixed-time nonovershooting stabilizers and safety filters by homogeneous feedback,''
{\em IEEE Transactions on Automatic Control},
vol.~68, no.~11, pp.~6434--6449, 2023.

\bibitem{Morse_1983}
A.~S.~Morse,
``Recent problems in parameter adaptive control,''
{\em Outils et Mod\`eles Math\'ematiques pour l'Automatique, l'Analyse de Syst\`emes et le Traitement du Signal},
vol.~3, pp.~733--740, 1983.


\bibitem{Nussbaum_1983}
R.~D.~Nussbaum,
``Some remarks on a conjecture in parameter adaptive control,''
{\em Systems \& Control Letters},
vol.~3, no.~5, pp.~243--246, 1983.

\bibitem{Xudong_tac_1998}
X.~Ye and J.~Jiang,
``Adaptive nonlinear design without a priori knowledge of control directions,''
{\em IEEE Transactions on Automatic Control},
vol.~43, no.~11, pp.~1617--1621, 1998.

\bibitem{ge_scl_2008}
S.~S.~Ge, C.~Yang, and T.~H.~Lee,
``Adaptive robust control of a class of nonlinear strict-feedback discrete-time systems with unknown control directions,''
{\em Systems \& Control Letters},
vol.~57, no.~11, pp.~888--895, 2008.

\bibitem{chenzy_auto_2019}
Z.~Chen,
``Nussbaum functions in adaptive control with time-varying unknown control coefficients,''
{\em Automatica},
vol.~102, pp.~72--79, 2019.

\bibitem{Scheinker2013tac}
A.~Scheinker and M.~Krsti\'c,
``Minimum-seeking for CLFs: Universal semiglobally stabilizing feedback under unknown control directions,''
{\em IEEE Transactions on Automatic Control},
vol.~58, no.~5, pp.~1107--1122, 2013.


\bibitem{Gurvits1992ICRA}
L.~Gurvits,
``Averaging approach to nonholonomic motion planning,''
in {\em Proceedings of the 1992 IEEE International Conference on Robotics and Automation},
pp.~2541--2542, IEEE, 1992.

\bibitem{Leblanc_1922}
M.~Leblanc,
``Sur l\'electrification des chemins de fer au moyen de courants alternatifs de fr\'equence \'elev\'ee,''
{\em Revue G\'en\'erale de l\'Electricit\'e},
vol.~12, no.~8, pp.~275--277, 1922.

\bibitem{Krs_es_2000}
M.~Krsti\'c and H.~H.~Wang,
``Stability of extremum seeking feedback for general nonlinear dynamic systems,''
{\em Automatica},
vol.~36, no.~4, pp.~595--601, 2000.


\bibitem{Stankovic_cdc_2010}
M.~S.~Stankovic, K.~H.~Johansson, and D.~M.~Stipanovic,
``Distributed seeking of Nash equilibria in mobile sensor networks,''
in {\em 49th IEEE Conference on Decision and Control},
pp.~5598--5603, IEEE, 2010.

\bibitem{Stankovic_auto_2010}
M.~S.~Stankovic and D.~M.~Stipanovic,
``Extremum seeking under stochastic noise and applications to mobile sensors,''
{\em Automatica},
vol.~46, no.~8, pp.~1243--1251, 2010.

\bibitem{Oliveira_tac_2020}
T.~R.~Oliveira, J.~Feiling, S.~Koga, and M.~Krsti\'c,
``Multivariable extremum seeking for PDE dynamic systems,''
{\em IEEE Transactions on Automatic Control},
vol.~65, no.~11, pp.~4949--4956, 2020.

\bibitem{Williams_2025ESSafe}
A.~Williams, M.~Krsti\'c, and A.~Scheinker,
``Semiglobal safety-filtered extremum seeking with unknown CBFs,''
{\em IEEE Transactions on Automatic Control},
vol.~70, no.~3, pp.~1698--1713, 2025.


\bibitem{Moreau2000tac}
L.~Moreau and D.~Aeyels,
``Practical stability and stabilization,''
{\em IEEE Transactions on Automatic Control},
vol.~45, no.~8, pp.~1554--1558, 2000.

\bibitem{Durr2013auto}
H.~B.~D\"urr, M.~S.~Stankovic, C.~Ebenbauer, and K.~H.~Johansson,
``Lie bracket approximation of extremum seeking systems,''
{\em Automatica},
vol.~49, no.~6, pp.~1538--1552, 2013.

\bibitem{Scheinker2014scl}
A.~Scheinker and M.~Krsti\'c,
``Extremum seeking with bounded update rates,''
{\em Systems \& Control Letters},
vol.~63, pp.~25--31, 2014.

\bibitem{Wang_tac_2024}
S.~Wang, M.~Guay, and R.~D.~Braatz,
``Extremum-seeking regulator for a class of nonlinear systems with unknown control direction,''
{\em IEEE Transactions on Automatic Control},
vol.~69, no.~12, pp.~8931--8937, 2024.

\bibitem{Scheinker_cdc_2012}
A.~Scheinker and M.~Krsti\'c,
``Extremum seeking-based tracking for unknown systems with unknown control directions,''
in {\em 51st IEEE Conference on Decision and Control},
pp.~6065--6070, IEEE, 2012.


\bibitem{Krstic1995backstepping}
M.~Krsti\'c, P.~V.~Kokotovi\'c, and I.~Kanellakopoulos,
{\em Nonlinear and Adaptive Control Design}.
New York, NY, USA: John Wiley \& Sons, Inc., 1995.

\bibitem{Allgower_2007}
P.~Wieland and F.~Allg\"ower,
``Constructive safety using control barrier functions,''
in {\em IFAC Proceedings Volumes},
vol.~40, no.~12, pp.~462--467, 2007.

\bibitem{Akametalu_2018}
J.~F.~Fisac, A.~K.~Akametalu, M.~N.~Zeilinger, S.~Kaynama, J.~Gillula, and C.~J.~Tomlin,
``A general safety framework for learning-based control in uncertain robotic systems,''
{\em IEEE Transactions on Automatic Control},
vol.~64, no.~7, pp.~2737--2752, 2018.

\bibitem{Mudgett_tac_1985}
D.~Mudgett and A.~Morse,
``Adaptive stabilization of linear systems with unknown high-frequency gains,''
{\em IEEE Transactions on Automatic Control},
vol.~30, no.~6, pp.~549--554, 1985.
\end{thebibliography}

\section*{Appendix}

\renewcommand{\theequation}{A.\arabic{equation}}
\setcounter{equation}{0}
\noindent
\textbf{Proof of Lemma 1.} Notice that system (\ref{sys_mm_orgin}) is GUUB, that is, there exists $\delta>0$ such that for all $t_0\in\R$, the following conditions are satisfied.

\noindent
$\bullet$ For every $c_2\in(\delta,\infty)$, there exists $c_1\in(\delta,\infty)$ such that
\begin{align}\label{ori_stable}
    |x_0|<c_1 \;\Rightarrow\;|\psi(t,t_0,x_0)|<c_2, \quad \forall t\in[t_0,\infty)
\end{align}
$\bullet$ For every $c_1\in(0,\infty)$, there exists $c_2\in(\delta,\infty)$ such that
\begin{align}\label{ori_bounded}
    |x_0|<c_1 \;\Rightarrow\;|\psi(t,t_0,x_0)|<c_2, \quad \forall t\in[t_0,\infty)
\end{align}
$\bullet$ For every $c_1, c_2\in(\delta,\infty)$, there exists $t_f\in(0,\infty)$ such that
\begin{align}\label{ori_attra}
    |x_0|<c_1 \;\Rightarrow\;|\psi(t,t_0,x_0)|<c_2, \quad \forall t\in[t_0+t_f,\infty)
\end{align}
We prove the conditions (\ref{uniform_stable})-(\ref{uniform_attractivity}) for system (\ref{sys_mm_ocia}).

\medskip
1) Take an arbitrary $c_2\in(\delta,\infty)$ and let $b_2\in(\delta,c_2)$. By (\ref{ori_stable}), there exists $c_1\in(\delta,\infty)$ such that
\begin{align}\label{uniform_stable_appen}
         |x_0|<c_1 \;\Rightarrow\; |\psi(t,t_0,x_0)|<b_2, \quad \forall t\in[t_0,\infty)
\end{align}
Let $b_1\in(\delta,c_1)$. By (\ref{ori_attra}), there exists $t_f\in(0,\infty)$ such that for all $t_0\in\R$
\begin{align}\label{uniform_attractivity_appen}
            |x_0|<c_1 \;\Rightarrow\; |\psi(t,t_0,x_0)|<b_1, \quad \forall t\in[t_0+t_f,\infty)
\end{align}
Let $d=\min\{c_1-b_1, c_2-b_2\}$. By the converging trajectories property and with $\mathcal{K}=\{\chi\in\R^n|\; |\chi|<c_1\}$, there exists $\bar{\epsilon}$ such that for all $t_0\in\R$ and for all $\epsilon\in(0,\bar{\epsilon})$
\begin{align}\label{conver_tra_pro_appen}
    & |x_0|<c_1 \;\Rightarrow\; |\psi^\epsilon(t,t_0,x_0)-\psi(t,t_0,x_0)|<d, \nonumber \\
    & \quad \quad
    \quad \quad
    \quad \quad
    \quad \quad
    \quad \quad
    \quad \quad
    \quad \quad \forall t\in[t_0,t_0+t_f]
\end{align}
By (\ref{uniform_stable_appen})-(\ref{conver_tra_pro_appen}), we have
\begin{align}\label{property1}
    |x_0|<c_1 \;\Rightarrow\;
        \left\{
            \begin{array}{ll}
            |\psi^\epsilon(t,t_0,x_0)|<c_2, &\; \forall t\in[t_0,t_0+t_f] \\
            |\psi^\epsilon(t,t_0,x_0)|<c_1, &\; \textup{for} \;\; t=t_0+t_f
            \end{array}
        \right.
\end{align}
Since $|\psi^\epsilon(t_0+t_f,t_0,x_0)|<c_1$, a repeated application of (\ref{property1}) gives that
\begin{align}\label{uniform_stable_ap_result}
         |x_0|<c_1 \;\Rightarrow\; |\psi^\epsilon(t,t_0,x_0)|<c_2, \quad \forall t\in[t_0,\infty)
\end{align}
which is (\ref{uniform_stable}).

\medskip
2) Take an arbitrary $c_3\in(\delta,\infty)$. By (\ref{ori_bounded}), there exists $b_2\in(\delta,\infty)$ for all $t_0\in\R$
\begin{align}\label{pro2_bound_appen}
         |x_0|<c_3 \;\Rightarrow\; |\psi(t,t_0,x_0)|<b_2, \quad \forall t\in[t_0,\infty)
\end{align}
Choose $b_3\in(\delta,c_3)$. By (\ref{ori_attra}), there exists $t_f\in(0,\infty)$ such that for all $t_0\in\R$
\begin{align}\label{property2}
    |x_0|<c_3 \;\Rightarrow\;|\psi(t,t_0,x_0)|<b_3, \quad \forall t\in[t_0+t_f,\infty]
\end{align}
Let $c_2\in(b_2,\infty)$, $d=\min\{c_3-b_3, c_2-b_2\}$. By the converging trajectories property and with $\mathcal{K}=\{\chi\in\R^n|\; |\chi|<c_3\}$, there exists $\bar{\epsilon}$ such that for all $t_0\in\R$ and for all $\epsilon\in(0,\bar{\epsilon})$
\begin{align}\label{conver_tra_pro2_appen}
    & |x_0|<c_3 \;\Rightarrow\; |\psi^\epsilon(t,t_0,x_0)-\psi(t,t_0,x_0)|<d, \nonumber \\
    & \quad \quad
    \quad \quad
    \quad \quad
    \quad \quad
    \quad \quad
    \quad \quad
    \quad \quad \forall t\in[t_0,t_0+t_f]
\end{align}
Combining (\ref{pro2_bound_appen})-(\ref{conver_tra_pro2_appen}), we have
\begin{align}\label{property2}
    |x_0|<c_3 \;\Rightarrow\;
        \left\{
            \begin{array}{ll}
            |\psi^\epsilon(t,t_0,x_0)|<c_2, &\; \forall t\in[t_0,t_0+t_f] \\
            |\psi^\epsilon(t,t_0,x_0)|<c_3, &\; \textup{for} \;\; t=t_0+t_f
            \end{array}
        \right.
\end{align}
Repeating the argument of (\ref{property2}) gives that
\begin{align}\label{property2_peat}
    |x_0|<c_3 \;\Rightarrow\;|\psi^\epsilon(t,t_0,x_0)|<c_2, &\quad \forall t\in[t_0,\infty]
\end{align}
By (\ref{property2_peat}), it gives that (\ref{uniform_bounded}) for every $c_1\in(\delta,\infty)$. The remaining is to prove that for every $c_1\in(0,\delta]$, (\ref{uniform_bounded}) still holds. The proof is straightforward  because for $c_1\in(0,\delta]$
\begin{align}
    |x_0|<c_1\; \Rightarrow\; |x_0|<c_3\;\Rightarrow\;|\psi^\epsilon(t,t_0,x_0)|<c_2, &\;\; \forall t\in[t_0,\infty]
\end{align}
which is (\ref{uniform_bounded}).

\medskip
3) Take arbitrary $c_1, c_2\in(\delta,\infty)$. By (\ref{uniform_stable_ap_result}), there exists $c_3\in(\delta,\infty)$ and $\epsilon\in(0,\epsilon_1]$
\begin{align}\label{uniform_stable_ap_result3}
         |x_0|<c_3 \;\Rightarrow\; |\psi^\epsilon(t,t_0,x_0)|<c_2, \quad \forall t\in[t_0,\infty)
\end{align}
Let $b_3\in(\delta,c_3)$. By (\ref{ori_attra}), there exists $t_f\in(0,\infty)$ such that
\begin{align}\label{uniform_attra}
         |x_0|<c_1 \;\Rightarrow\; |\psi(t,t_0,x_0)|<b_3, \quad \forall t\in[t_0+t_f,\infty)
\end{align}
Let $d=c_3-b_3$. By the converging trajectories property and with $\mathcal{K}=\{\chi\in\R^n|\; |\chi|<c_1\}$, there exists $\bar{\epsilon}$ such that for all $t_0\in\R$ and for all $\epsilon\in(0,\bar{\epsilon})$
\begin{align}\label{conver_tra_pro3_appen}
    & |x_0|<c_1 \;\Rightarrow\; |\psi^\epsilon(t,t_0,x_0)-\psi(t,t_0,x_0)|<d, \nonumber \\
    & \quad \quad
    \quad \quad
    \quad \quad
    \quad \quad
    \quad \quad
    \quad \quad
    \quad \quad \forall t\in[t_0,t_0+t_f]
\end{align}
Combining (\ref{uniform_attra}) and (\ref{conver_tra_pro3_appen}), we have
\begin{align}\label{pro3}
         |x_0|<c_1 \;\Rightarrow\; |\psi^\epsilon(t,t_0,x_0)|<c_3, \quad \textup{for} \;\; t=t_0+t_f
\end{align}
Together with (\ref{uniform_stable_ap_result3}), we have for all $\epsilon\in(0,\min\{\epsilon_1,\bar{\epsilon}\})$
\begin{align}\label{pro3}
         |x_0|<c_1 \;\Rightarrow\; |\psi^\epsilon(t,t_0,x_0)|<c_2, \quad \forall  t\in[t_0+t_f,\infty)
\end{align}
which is (\ref{uniform_attractivity}).

\end{document}